\documentclass[11pt,twoside]{article}
\usepackage{rjr-asp}
\usepackage{lscape}
\usepackage{epsfig,graphicx,natbib,url}  
\begin{document}

\markboth{T. Wiegelmann}{Preprocessing of Hinode/SOT data}
\title{Preprocessing of Hinode/SOT vector magnetograms for nonlinear
force-free coronal magnetic field modelling.}
\author{T. Wiegelmann $^{1}$,
        J.K. Thalmann $^1$,
        C.J. Schrijver $^{2}$,
        M.L. DeRosa $^{2}$ and
        T.R. Metcalf $^3$}
\affil{$^{1}$ Max-Planck-Institut f\"ur Sonnensystemforschung,
Max-Planck-Strasse 2, 37191 Katlenburg-Lindau, Germany \\
$^{2}$ Lockheed Martin Advanced Technology Center, Dept. ADBS. Bldg. 252,
              3251 Hanover St., Palo Alto, CA 94304, USA \\
 $^{3}$ Northwest Research Associates, Colorado Research Associates Division,
              3380 Mitchel Ln, Boulder, CO 90301 USA}

\begin{abstract}
 The solar magnetic field is key to understanding the physical processes in the solar atmosphere.
 Nonlinear force-free codes  have been shown to
 be useful in extrapolating the coronal field from
 underlying vector boundary data \citep[see][for an overview]{Schrijver:etal06}.
 However, we can only measure the magnetic field vector routinely
 with high accuracy in the photosphere with, e.g., Hinode/SOT, and unfortunately these data do not
 fulfill the force-free consistency condition as defined by \cite{aly89}. We must therefore apply
 some transformations to these data before nonlinear force-free extrapolation codes can be legitimately
 applied.  To this end, we have developed
 a minimization procedure  that uses the measured photospheric field vectors as input to approximate a
 more chromospheric like field \citep[The method was dubbed preprocessing. See][for details]{wiegelmann:etal06}.
 The procedure includes force-free consistency integrals and spatial smoothing.
 The method has been intensively tested with model active regions \citep[see][]{metcalf:etal07} and been applied
 to several ground based vector magnetogram data before. Here we apply the
 preprocessing program to photospheric magnetic field measurements with the Hinode/SOT instrument.
\end{abstract}

\begin{figure}  
  \centering
\includegraphics[height=8cm]{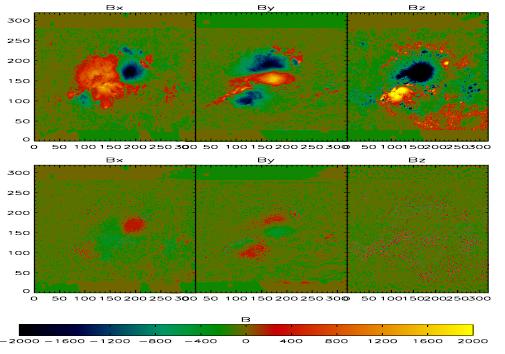}
\caption[]{\label{figure1}  
Top: Photospheric vector magnetogram data measured with the Hinode/SOT instrument
for a flaring active region, december 12, 2006. Bottom: Difference between
the original and the preprocessed data.
The preprocessed dataset provides us suitable boundary conditions for a nonlinear
force-free extrapolation of the coronal magnetic field.
}\end{figure}

\section{Results}      \label{results}
The original Hinode vector magnetogram ($B_x, B_y, B_z$ components of
the photospheric magnetic field vector, respectively) is shown in the top panel of
Fig. \ref{figure1}. The solar photosphere has a plasma $\beta$ of about unity and
non-magnetic forces like pressure gradients and gravity are important here.
Consistent boundary conditions for a nonlinear force-free extrapolation of the
coronal magnetic field require, however, that the net force on the boundary vanishes.
We introduce a dimensionless number which measures the net forces, normalized by the photospheric
magnetic field strength:
$$
\epsilon_{\mbox{force}}=\frac{|\int_{S} B_x B_z \;dx\,dy| + |\int_{S} B_y B_z \;dx\,dy|+
|\int_{S} (B_x^2+B_y^2)-B_z^2 \;dx\,dy |}
{\int_{S} (B_x^2+B_y^2+B_z^2) \;dx\,dy},
$$
and only for $\epsilon_{\mbox{force}} \ll 1$ the boundary conditions are consistent with a nonlinear
force-free coronal magnetic field. We find $\epsilon_{\mbox{force}}=0.16$ for the original
photospheric Hinode magnetogram. This is significantly better compared to ground based observations,
where one finds values up to about unity, but still not sufficient in order to use the
data as boundary conditions for a force-free magnetic field extrapolation directly. After
applying the preprocessing procedure  we find
$\epsilon_{\mbox{force}}=1.8 \cdot 10^{-4}$, almost three orders of magnitude lower
than the original data. This preprocessed vector magnetogram (and also the one of the following day)
has been used as boundary condition for nonlinear force-free computations of the coronal magnetic field before
and after an X-class flare (Schrijver et al., ApJ in preparation).

\acknowledgements We acknowledge the use of data from Hinode/SOT. TW was supported by
DLR, JT by DFG and CS, MD, TM by Lockheed Martin independent research funds.
We dedicate this work to our colleague Tom Metcalf, who died in July 2007.







\end{document}